# The Effect of Object-Oriented Programming Expertise in Several Dimensions of Comprehension Strategies

Jean-Marie Burkhardt*,
Françoise Détienne*
Susan Wiedenbeck**

* EIFFEL, Cognition and Cooperation in Design, INRIA
Domaine de Voluceau, Rocquencourt, BP 105,
78153, Le Chesnay, cedex, France
Jean-Marie.Burkhardt@inria.fr, Francoise.Detienne@inria.fr

** Computer Science and Engineering Department,
University of Nebraska
Lincoln, NE 68588-0115, USA
susan@cse.unl.edu

**Abstract**

This study analyzes object-oriented (OO) program comprehension by experts and novices. We examine the effect of expertise in three dimensions of comprehension strategies: the scope of the comprehension, the top-down versus bottom-up direction of the processes, and the guidance of the comprehension activity.
Overall, subjects were similar in the scope of their comprehension, although the experts tended to consult more files. We found strong evidence of top-down, inference-driven behaviors, as well as multiple guidance in expert comprehension. We also found evidence of execution-based guidance and less use of top-down processes in novice comprehension. Guidance by inheritance and composition relationships in the OO program was not dominant, but nevertheless played a substantial role in expert program comprehension. However, these static relationships more closely tied to the OO nature of the program were exploited poorly by novices.
To conclude, these results are discussed with respect to the literature on procedural program comprehension.

## 1. Introduction

This study analyzes object-oriented (OO) program comprehension by experts and novices. The object-oriented paradigm is growing fast in popularity, but not enough scientific evidence has been amassed about it. The research that exists is mostly focused on program design and reuse (see for example [5; 4; 11]). Furthermore, there is little empirical work on the comprehension processes of OO programmers. Most previous studies on the comprehension of software texts were carried out in the context of procedural or declarative languages.

The strategies invoved in program comprehension may be characterized along three dimensions: the scope of the comprehension, the top-down versus bottom-up direction of the processes, and the guidance of the comprehension activity. Our objective is to analyze OO program comprehension and to examine the effect of expertise in these three dimensions of comprehension.

## 2. Objectives

The scope of the comprehension refers to the quantity and nature of information contained in a program which is actually read and processed when understanding that program. Information processing is highly selective, in particular when the quantity of information to process is high and when only a subset of information is relevant for the task at hand. Whereas the scope of comprehension has been analysed with respect to the task factor [9, 8], the dimensions of quantity and nature of information have often been neglected in comprehension studies because of the small size of the programs used.

The top-down (or knowledge-driven) versus bottom-up (or data-driven) direction of the processes refers to the use of problem and programming domain knowledge (called programming plans or schemas) in program comprehension. Numerous studies have shown that such knowledge is activated and allows experts to draw inferences (e.g., expectation systems) while reading a program (see for example [3, 6, 12, 13]. Clearly, these studies show that there is an effect of expertise in this dimension:

comprehension processes of experts are more top-down whereas those of novices are more bottom-up.

The guidance of the comprehension refers to the nature of the representations constructed in program comprehension. Different kinds of abstraction may be constructed (e.g., data flow, control flow, function), and an important issue is to identify which kinds of relationships guide the comprehension process. This issue has been explored in the mental model approach to program comprehension contrasting the kind of information relationships making up the program model and the situation model [1, 10, 2]. Empirical studies show that the guidance of comprehension is affected by expertise [1] and by the task [7]. Bergantz and Hassell analysed the protocols of three programmers of different levels of experience understanding a program which solved a toy problem. They found that the least experienced programmer relied more on symbolic execution than on other other kinds of relationships. By contrast, experts used multiple guidance.

These studies have two limitations. First, most of them used very short programs. Short programs are clearly not sufficient for analyzing the scope dimension. As concerns the dimension of guidance, the kind of relationships examined in previous studies are at a rather low level of granularity, which is sufficient for short programs but not necessarily for larger programs. A second limitation comes from the procedural or declarative nature of the programming languages used in these studies. Guidance has been examined according to information relationships relevant to the procedural or declarative nature of the paradigms used. Clearly, in OO program comprehension, we need to account for the use of relationships (e.g. classes, inheritance, composition) which are more related to the OO nature of the paradigm.

Our objective is to analyze OO program comprehension and to examine the effect of expertise in the three dimensions of comprehension. One of our research questions is whether experts will show a different scope of comprehension than novices. With respect ot the direction of comprehension, we expected experts to use more top-down processes than novices. In an OO program this can be explored along three dimensions of abstraction: implementation level, inheritance hierarchy, and calling hierarchy. With respect to guidance of comprehension, we expected that experts would use multiple guidance more than novices. We also expected experts to use abstractions related to the OO paradigm for guidance, such as the relationships of classes. Novices were expected to use specifically OO abstractions less often.

The present paper reports on part of a larger study [2] which investigated the effect of a group of factors in OO program comprehension: expertise, phase of comprehension, and task. In this paper we report on the first phase of comprehension in which subjects were asked to read the program in order to later perform a task. We did not expect a strong effect of this task orientation. While our first concern is to analyze the effect of expertise, we make reference to task orientation when appropriate.

## 3. Methodology

A two-factor between subjects design was used. The factors were expertise (OO expert vs. OO novice) and task orientation (documentation vs. reuse orientation).

The subjects were 28 object-oriented experts and 21 object-oriented novices. The experts were professional programmers with experience in object-oriented design with C++. The novices were advance computer science students who were experienced in C but had only a basic knowledge of object-oriented programming and C++. Twenty-eight of the subjects were speakers of English and 21 were speakers of French.

The materials consisted of a database program of approximately 550 lines which managed personnel, student, and course information for a small university. The program was composed of 10 classes. The hierarchical organization of the classes is shown in figure 1. The program was written in object-oriented C++ and presented in 21 files. The domain of the problem allowed us to write a program which took good advantage of the OO paradigm, including ease of conceptualization in terms of objects, classes, and inheritance. As in past comprehension studies (Pennington, 1987a), little documentation was included in the text of the program.

**Figure 1 : Hierarchical tree of classes for the database program**

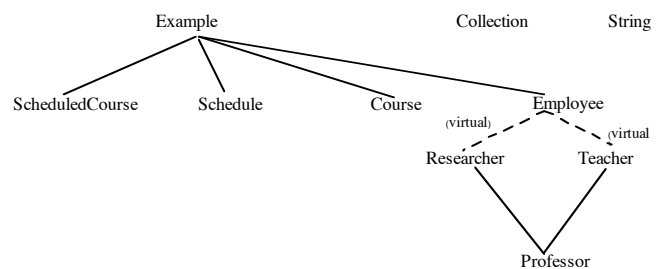

Experts and novices were assigned randomly to the documentation or reuse groups. They were given an orientation to study the program for later reuse or documentation, as appropriate. Subjects were then given the database program and asked to study it for 35 minutes. Verbal protocols were collected, and the files consulted as well as the transitions between files were recorded.

For the coding of protocols, we divided the 35 minutes spent in program comprehension into three equal stages of 11.66 minutes each. Each file consultation activity was characterized by the name of the file consulted, the implementation level of the file consulted, its level of abstraction in the inheritance hierarchy, and its level of abstraction in the calling hierarchy. Also, each transition between files was characterized according to the kind of relationship guiding the transition, as indicated by the verbal protocols: execution, includes, inheritance, composition, methods, variables, random, or other.

## 4. Results

### 4.1 Scope of Comprehension

In order to examine the scope of comprehension, an analysis of variance was performed on the mean number of different files consulted. The between subjects factors were expertise (novice or expert) and task orientation (documentation or reuse). The within subjects factors were stage (1, 2, 3) and the type of file (.h and .cc). There was no overall effect of task orientation or stage. Although the effect of the expertise was not significant, the scope of comprehension of the experts tended to be wider than that of novices ($m_{expert}$= 3.993, sd=2.304; $m_{novice}$=3.402, sd= 2.064; $F(1, 45)$= 3.825, $p<0.0567$). There was an overall significant effect of type of file (m.hfile= 4.207, sd=2.345; m.ccfile=3.272, sd= 1.988; $F(1, 45)$= 20.511, $p<0.0001$). The subjects consulted significantly more .h files than .cc files. The two-way and higher order interactions were not significant. See table 1 for the means of the interactions.

**Table 1 . Means (sd) for the number of .h files and .cc files consulted, by stage and by expertise**

|  | .h files | |
|---|---|---|
|  | Experts | Novices |
| stage 1 | 4.341 (2.903) | 4.029 (2.562) |
| stage 2 | 5.165 (1.940) | 4.176 (2.156) |
| stage 3 | 4.093 (2.020) | 3.114 (2.085) |
| global | 4.533 (2.346) | 3.773 (2.290) |
|  | .cc files | |
|  | Experts | Novices |
| stage 1 | 2.857 (2.24) | 3.476 (1.887) |
| stage 2 | 3.679 (2.038) | 2.857 (2.081) |
| stage 3 | 3.821 (2.091) | 2.762 (1.136) |
| global | 3.452 (2.142) | 3.032 (1.750) |

We also examined the nature of the file consulted. The results are shown in Table 2. We contrasted which files were consulted at least once by most subjects with which files were consulted at least once by only a few subjects. Globally the most consulted files were Prof.h (consulted by 48 out of 49, i.e. 98% of the subjects) and Main.cc (consulted by 47 out of 49 subjects, i.e. 96%). We can remark that Init.cc was the sixth most consulted file (consulted by 43 out of 49 subjects, i.e. 88%). The least consulted files were Makefile and Scheduled_course.cc, each of which was consulted by less than half the subjects (22 out of subjects, i.e. 47%).

**Table 2 : percentage of subjects consulting each file for the 35 minutes**
**(* indicates that there was no .cc file)**

|  | .h | .cc |
|---|---|---|
| includes | 80% | * |
| main | 57% | 96% |
| init | 78% | 88% |
| example | 94% | * |
| collection | 94% | 94% |
| scheduledcourse | 69% | 45% |
| schedule | 69% | 65% |
| course | 86% | 61% |
| employee | 88% | 49% |
| research | 84% | * |
| teacher | 92% | 69% |
| professor | 98% | 86% |
| string | 65% | 55% |
| makefile |  | 47% |

Considering that the .h files were globally more consulted than the .cc files, we repeated the same analysis as above for each type of file separately. This was done for the three stages together and separately. The results of this analysis for the three stages together are summarized graphically in Figure 2 which represent the hierarchical organization of the program classes.

**Figure 2 : most and least consulted .h and .cc files :"+" for most consulted files, "-" for least consulted files**

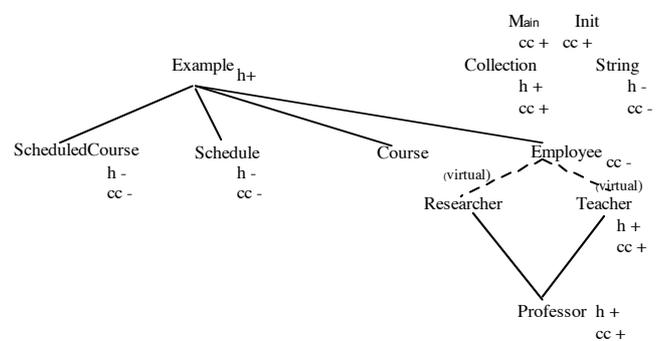

In Figure 2, it can be seen that the .h files which were consulted most by subjects are Example.h, Collection.h, Professor.h, and Teacher.h. The .cc files which were consulted most by subjects are Collection.cc, Professor.cc, and Teacher.cc. Clearly there is a correspondance between the consultation of .h files and .cc files. The classes for which the .h file were the most consulted are also the classes for which the .cc file were the most consulted[1]. The most consulted classes are at the top of the hierarchy (Example) and at the bottom of the hierarchy (Professor and Teacher). It should be noted that the classes at the bottom of the hierarchy are also the classes which are first referenced in the main function. The class Collection, which is not in the hierarchy of classes, was also among the most consulted. This is a class used by most other classes to perform most of the procedural processing.

In Figure 2, it can be seen that the .h files which were consulted by the smallest number of subjects are Scheduled_course.h, Schedule.h, and String.h. The .cc files which were consulted by the smallest number of subjects are: Scheduled_course.cc, Employee.cc, and String.cc. The classes for which the .h file were the least consulted correspond partially to the classes for which the .cc file were the least consulted. The least consulted classes are: (1) classes located either in the middle of the hierarchy (Scheduled_course, Schedule, Employee) and, (2) a class representing a familiar type of data (String), which is probably considered equivalent to a predefined class.

We also considered sparately the three stages of reading. An interesting result is that the most consulted classes are either at the top (Example) or at the bottom (Professor) of the hierarchy in stage 1. In stage 2 and stage 3, the most consulted classes can be found at those same places in the hierarchy and also following up the hierarchy (Teacher). We can interpret this last result in two ways. The programmers are guided by the structure of the notation. In the C++ notation there is a one way link from a class to its superclass. Another interpretation is related to the procedural links between these classes. In Main, Professor is the first class called and Teacher the second class called. Also, Professor makes many calls to methods defined in its superclass[2], Teacher. Thus, following this procedural view could explain why subjects follow up the hierarchy.

We also analysed the nature of the files consulted as a function of expertise. We found that each group behaves quite similarly to the whole population. There are several exceptions: 92 percent of experts in the documentation condition consulted the Teacher.cc file whereas 40 percent of novices in the reuse group consulted this file. In stage 1, collection.cc was consulted more by novices (81%) than experts (57%) and the init.cc file was consulted more by novices in the documentation condition (72%) than the other groups (RN: 40%; RE: 33%; DE: 30%).

## 4.2 Top-down versus bottom-up processes

The direction, top-down versus bottom-up, of the understanding processes can be examined according to the level of abstraction of the entities of the program which are read first. We have considered abstraction along three dimensions: (1) the class header file (.h file) is more abstract than the implementation file (.cc file) inasmuch as the header file contains the class declaration including delarations of attributes and methods, while the code implementation of methods is made in the .cc file, (2) the level of abstraction in the hierarchy of classes and, (3) the level of abstraction in the hierarchy of calls.

### 4.2.1 Header versus implementation

Our first working hypothesis is that reading .h files reflects top-down processes whereas reading .cc files reflects bottom-up processes. An analysis of variance was performed on the number of files consulted, separately for the .h files and the .cc files. The between subjects factors were expertise (novice or expert) and task orientation (documentation or reuse). The within subjects factor was stage (1, 2, 3). For the .h files, there was no overall effect of task orientation or stage. The effect of the expertise approached significance ($m_{expert}$= 4.533, sd=2.346; $m_{novice}$=3.773, sd= 2.290; $F(1, 45)$= 3.858, $p<0.0557$). This difference was significant for stage 2 and stage 3 together ($m_{expert}$= 4.629, sd=2.035; $m_{novice}$=3.645, sd=2.163; $F(1, 45)$=5.595, $p<0.0224$). Thus, the experts tend to read more .h files than the novices whatever the stage. These results could be interpreted as reflecting more top-down processes in the reading strategies of experts than in those of novices. For the .cc files, there was no overall effect of task orientation, expertise, or stage. The effect of the expertise was significant in stage 3 ($m_{expert}$= 3.821, sd=2.091; $m_{novice}$=2.762, sd=1.136; $F(1, 45)$= 4.445, $p<0.0406$). Thus the experts read more .cc files than novices in stage 3 only. Our interpretation of these results is that the experts gained an abstract view of the program through a top-down approach in earlier stages and later concentrate on implementation details.

A complementary analysis was conducted to examine the evolution of top-down versus bottom-up processes in each group. In each group, we compared the consultations of .h files versus .cc files in each stage (see table 1, previously given). We found that, the experts consulted significantly more .h files than .cc files in stage 1 ($m_{expert/.hfile/stage1}$= 4.341, sd=2.903; $m_{expert/.ccfile/stage1}$=2.857, sd=2.240; $t(27)$= -2.359, $p<0.0258$) and in

---

[1] Except for the Example class because this class has no .cc file.

[2] By contrast, Professor makes very little use of its other superclass, Researcher, which may explain why this class is less consulted than its sibling class, Teacher.

stage 2 (mexpert/.hfile/stage2= 5.165, sd=1.940; mexpert/.ccfile/stage2=3.679, sd=2.038, t(27)= -2.965, p<0.0063). In stage3, the experts consulted as many .h files as .cc files. We found that, the novices consulted significantly more .h files than .cc files in stage 2 (mnovice/.hfile/stage2= 4.176, sd=2.156; mnovice/.ccfile/stage2=2.857, sd=2.081; t(20)= -2.56, p<0.0187. In stage1 and stage3, the novices consult as many .h files as .cc files. So clearly, the strategies of reading of experts involve more top-down processes in the first two stages whereas the strategies of reading of novices involve more top-down processes only in the second stage.

### 4.2.2 Levels of abstraction in the hierarchy of classes

We can also contrast top-down versus bottom-up strategies along a dimension of abstraction corresponding to the hierarchy of classes. We have seen that globally the most consulted classes are at the top and the lowest level of the hierarchy and that this is observed whatever the expertise. This analysis does not show any dominance of top-down over bottom-up processes but rather that both processes are involved in comprehension by experts and novices. By examining the classes read in stage2, we find that the classes at the second level of the hierarchy are read by more experts than by novices (see Table 3). Thus experts seem to follow down the hierarchy of abstraction of classes.

**Table 3: percentage of the 2nd level classes consulted by the novices vs the experts during stage 2.**

|         | novices | experts |
|---------|---------|---------|
| classes | 60%     | 89%     |
| .h      | 40%     | 59%     |
| .cc     | 19%     | 30%     |

### 4.2.3 Levels of abstraction in the hierarchy of calls

We have also contrasted top-down versus bottom-up strategies along a dimension of abstraction corresponding to the hierarchy of calls. The level of granularity of our analysis does not allow us to analyse the order in which methods were read. However, we were able to analyse whether Main, which is at the top of the hierarchy of calls, was read early in the comprehension activity. We found that in stage 1, Main tended to be read more by experts than by novices (for main.h: 50% of experts and 38% of novices; for main.cc: 85% of experts and 62% of novices). However these differences are not significant.

### 4.3 Guidance of Comprehension

The third dimension which we use to characterize the reading/comprehension strategies of our subjects is the guidance of the comprehension process. We can distinguish several different approaches used to guide reading and comprehension. These are defined below.

- **Execution**: In this approach the comprehender follows the order of execution of the program while reading. Typically, the comprehender begins with the main function and follows the calls more or less systematically to other functions and methods in the program. This guidance of comprehension corresponds to a mental execution of the program.
- **Includes:** An includes approach was used when a comprehender read the program by following the order of the inculdes statements embedded in files. The comprehender usually began with main, then proceeded to read the first file named in the includes statements in main, then followed the includes in that file, etc., backing up when the end of a path of includes was reached. Thus, this type of guidance tended to take on the quality of a depth-first search.
- **Relationship among classes:** This approach involves reading the program in a manner which highlights the relationships among different classes. Two specific types of guidance are possible reflecting inheritance and composition relationships in a program.
  - **Inheritance:** In this approach the comprehender reads the program by following inheritance relationships among classes. For example, the comprehender may determine which class is the top of the hierarchy of classes and then move from that class to its subclasses, and so on down the inheritance hierarchy. Alternatively, the comprehender might begin at a class lower in the hierarchy and follow the inheritance relationships upwards.
  - **Composition**: Composition relationships occur when a class contains attributes (i.e., data members in C++) which are members of another class. For example, in our experimental program the class Course contained an attribute called offerings which was a member of the class Collection. An approach of reading guided by composition relationships would be indicated by a shift of attention from one class to another when such an attribute was encountered while reading.
- **Methods:** This approach involves reading a program to understand the internal workings of a class. Using this type of guidance, the comprehender attempts to find information about a method or to systematically follow the code of a method.
- **Variables:** In this strategy the comprehender attempts to find information about a variable or to follow the use of a variable. This strategy is largely internal to a class because it involves tracing the flow of data through the class. However, it may also cross class boundaries if a variable is passed as an argument to a method of another class.
- **Random**: The comprehender who uses this approach does not read according to information relationships in the program but rather in a manner which is effectively arbitrary with respect to information relationships. In our experiment, subjects were considered to be using a random strategy when they read the program in the order of the physical hardcopy of files or, equivalently, the order of the on-line listing of files. In the physical hardcopy and the on-line listing the files were in alphabetical order by file name, so the order did not represent meaningful relationships among files.
- **Other**: This designation was used when the approach being used by a comprehender could not be identified from the combination of physical activities and verbalizations recorded.

### 4.3.1 Global Guidance

Our first analysis of the guidance of comprehension of the program concerned the type of global guidance used by subjects[3]. We determined for each subject whether a dominant global type of guidance of reading could be identified. As an operational definition, a subject was considered to have a dominant global guidance type if the subject used a given type 30 percent or more of the time in two stages and 30 percent or more of the time in the three stages together (i.e., in the total 35 minutes). By these criteria, it was possible for a subject to have more than one dominant global strategy, and this occurred in four cases. We found that a dominant global strategy could be identified in 80 percent of the subjects (38 out of 49), including 86 percent of the novices and 75 percent of the experts. The distribution of the global strategies is shown by expertise in Table 5. As the table suggests, three strategies predominated: random, execution, and methods. These strategies predominated irrespective of task orientation. Outside of these three strategies, few other global strategies were observed. Seven percent of experts (2 subjects) used predominantly the inheritance approach, and 5 percent of novices (1 subject) used an indeterminant strategy. Interestingly, one quarter of the experts did not have a predominant global strategy by our criteria.

---
[3] The small amount of data prevent us from using statistical analysis in this part of the paper.

**Table 5. Percentage of subjects using each type of global guidance**

| Strategy | Novice | Expert |
|---|---|---|
| Execution | 29 | 7 |
| Includes | 0 | 0 |
| Methods | 10 | 14 |
| Inheritance | 0 | 7 |
| Composition | 0 | 0 |
| Random | 57 | 50 |
| Variables | 0 | 0 |
| Other | 5 | 0 |
| None | 14 | 25 |

### 4.3.2 Local Guidance

For a finer grained view of the guidance of comprehension, we also analyzed the local guidance used in each of the three stages of the comprehension period. A subject was classified as using a given type of local guidance if the subject used the type 30 percent or more of the time during a stage. It should be noted that by this criterion a subject could have more than one type of local guidance in a stage. The local guidance is shown by type for novices and experts in Table 6. A dominant local guidance type could be identified among experts and novices in similar proportions. There was a mean of 3.3 dominant types across the 3 stages in experts and 3.6 in novices. These dominant types are not necessarily distinct, since a subject often used the same dominant guidance type in more than one stage. There were few cases in which a dominant local guidance type could not be identified, and these decreased over time, 6 in stage 1, 3 in stage 2, and 1 in stage 3. Interestingly, the 6 subjects without a dominant local guidance type in stage 1 were all experts. In stage 2, only 2 experts did not have a dominant type, and in stage 3 the one subject without a dominant type was an expert.

We also determined the breadth of subjects' repertoire of guidance approaches by analyzing the number of distinct approaches used by each subject. The number of approaches exercised was somewhat larger in experts than in novices. Leaving out the random strategy, which can be considered a default, experts used a mean of 1.78 distinct approaches over the three stages and novices used a mean of 1.38 distinct approaches. However, this difference is not statistically significant. An examination of Table 6 shows that, while the three dominant global approaches were random, execution, and methods, additional approaches were used for local guidance. These included all of the other approaches defined above: inheritance, composition, includes, and variables. The random approach was still most common regardless of expertise and task; however, its use decreased over time, from 57 percent across all subjects in phase 1, to 47 percent in phase 2, to 27 percent in phase 3. This weak approach tended to be more used by novices than experts in the first two stages but its use was equal in the third stage. In terms of the consistency of approaches, the same approach was used across all three stages by 25 percent of the subjects (6 novices and 6 experts). In 10 of these cases the consistent approach was random (6 novices and 4 experts), and in 2 cases, both experts, it was methods.

**Table 6. Percentage of subjects using each type of local guidance**

| | Novice | | |
|---|---|---|---|
| Strategy | Stage 1 | Stage 2 | Stage 3 |
| Execution | 19 | 24 | 33 |
| Includes | 5 | 5 | 0 |
| Inheritance | 10 | 10 | 10 |
| Compos | 10 | 5 | 0 |
| Methods | 10 | 5 | 33 |
| Variables | 0 | 0 | 10 |
| Random | 71 | 57 | 29 |
| Other | 5 | 5 | 14 |
| None | 0 | 5 | 0 |
| | Expert | | |
| Strategy | Stage 1 | Stage 2 | Stage 3 |
| Execution | 14 | 11 | 7 |
| Includes | 7 | 4 | 4 |
| Inheritance | 4 | 25 | 14 |
| Compos | 7 | 7 | 4 |
| Methods | 7 | 21 | 46 |
| Variables | 4 | 7 | 11 |
| Random | 46 | 43 | 29 |
| Other | 0 | 4 | 0 |
| None | 21 | 7 | 4 |

### 4.3.3 Dynamic Versus Static Guidance

It is interesting to observe the use of guidance focusing on dynamic vs. static relationships in the program at both the global and the local level. Execution guidance can be considered dynamic because it involves an execution order trace. Inheritance and composition guidance can be considered static because they focus on fixed relationships of objects defined in the program. At the level of global guidance, 29 percent of novices used dynamic execution order guidance, but no novices used static guidance. For experts, only 7 percent used execution order guidance globally, while 7 percent used static inheritance order guidance globally. The use of dynamic and static local guidance is shown in Tables 7a and 7b.

**Table 7a. Number of subjects using guidance involving a dynamic view of the program (execution)**

| | Reuse | Documentation | |
|---|---|---|---|
| Novice | 5/10 | 6/11 | 52% |
| Expert | 5/15 | 3/13 | 29% |
| | 40% | 38% | |

**Table 7b. Percentage of subjects using guidance involving a static view of the program (inheritance or composition)**

| | Reuse | Documentation | |
|---|---|---|---|
| Novice | 4/10 | 2/11 | 29% |
| Expert | 9/15 | 3/13 | 43% |
| | 52% | 21% | |

As can be seen, novices tended to use dynamic execution order guidance locally more than experts, but there were no differences based on task orientation. The static inheritance and composition order guidance types were used locally more by experts than by novices, and there also appears to be a difference based on task, with the reuse oriented group using this view locally more than the documentation oriented group. An interpretation of the higher use of static guidance in reuse is the following. The subjects with a reuse orientation may systematically trace inheritance relationships because reuse in OO programs is done through the inheritance mechanism, i.e., reusing existing classes by adding new subclasses to specialize them to the specifications of the reuse problem. Reuse by inheritance is more likely to be understood and used by experts, and our results show that it is mostly experts in reuse who use static guidance (9/15).

## 5. Discussion and Conclusion

Our results can be summarized in terms of scope, direction, and guidance of comprehension strategies. There was a wide scope of comprehension as shown by the consultation of files. While not all files were consulted by all subjects, generally subjects consulted most of the files. Overall, subjects were similar in the scope of their comprehension, although there was a trend for experts to consult more files. Both experts and novices tended to consult highly files at the top of the hierarchy of classes and also files which were salient from a procedural perspective (i.e., files in which much of the procedural processing of data elements was done).

As expected, the direction of comprehension strategies of expert subjects was mostly top-down. This top-down direction was seen in two of the three dimension of their comprehension behavior. First, from the beginning of the comprehension period, they consulted abstract header files more than detailed implementation files. Second, in terms of the abstraction corresponding to the hierarchy of classes, experts read files at both the top and bottom of the hierarchy early in the comprehension activity, showing both top-down and bottom-up direction. However, in the middle stage of reading they appear to have followed down the hierarchy of classes. For novices, comprehension was not clearly top-down from the beginning. Novices tended to read the abstract header files less than experts overall and, unlike the experts, did not focus on them in the earliest stage of comprehension. Novices consulted classes at the top and bottom of the hierarchy of classes but did not show the tendency to follow down the hierarchy to middle levels as comprehension progressed.

In terms of the guidance of comprehension strategies, experts exercised multiple guidance, as expected. This is seen through the variety of local guidance methods used. It is also seen through the substantial number of experts with no globally dominant guidance approach, which suggests that they flexibly used different methods. It is also notable that experts used both dynamic and static guidance methods for local guidance. Novices showed less evidence of multiple guidance. As expected, their approach was mostly dynamic execution-based guidance with less use of static approaches.

Broadly, our results were similar to those previously found in procedural and declarative languages [1, 3, 6, 12, 13]. Like the earlier studies, we found strong evidence of top-down, inference-driven behaviors, as well as multiple guidance in expert comprehension. We also found evidence of execution-based guidance and less use of top-down processes in novice comprehension. We found that guidance by inheritance and composition relationships in the OO program was not dominant, but nevertheless played a substantial role in expert program comprehension. It is clear that various types of guidance are important in OO program comprehension, both static and dynamic types. However, the static relationships more closely tied to the OO nature of the program were exploited poorly by novices. We attribute this to their lack of experience with the OO paradigm.

Generalization of our results must be limited by the fact that they are based on the analysis of the comprehension of only one OO program.